\documentclass[12pt,showkeys,showpacs,] {revtex4}
\usepackage{graphicx}
\begin{document}

\title[Short Title]{Multi-Receiver Quantum Dense Coding with Non-Symmetric Quantum Channel}

\author{Chang-Bao \surname{FU}}
\author{Shou \surname{ZHANG}\footnote{E-mail:
szhang@ybu.edu.cn} } \affiliation{Department of Physics, College of
Science, Yanbian University, Yanji, Jilin 133002, PR China}

\author{Ke \surname{LI}}
\affiliation{Purchasing Office, Yanbian University, Yanji, Jilin
133002, PR China}

\author{Kyu-Hwang \surname{YEON}}
\affiliation{Department of Physics,  Institute for Basic  Science
Research, College of Natural Science, Chungbuk National University,
Cheonju, Chungbuk 361-763}

\author{Chung-In \surname{UM}}
\affiliation{Department of Physics, College of Science, Korea
University, Seoul 136-701}

\begin{abstract} A two-receiver quantum dense coding scheme and
an $N$-receiver quantum dense coding scheme, in the case of non-symmetric
Hilbert spaces of the particles of the quantum channel, are
investigated in this paper. A sender can send his messages to many
receivers simultaneously. The scheme can be applied to quantum
secret sharing and controlled quantum dense coding.
\end{abstract}



\maketitle The quantum entanglement among the quantum systems can
be used to perform many tasks, such as quantum cryptography
\cite{0001,0002}, quantum secret sharing
\cite{0003,0004,0005,0006} and so on. Quantum dense coding is also
one of the applications of entanglement in quantum communication.
Since Bennet and Wiesner \cite{0007} first proposed the quantum
dense coding (QDC) scheme, different QDC schemes have been
presented. For example, Lee {\it et al.} \cite{0008} have studied
QDC scheme among multiparties. Bose {\it et al.} \cite{0009} have
studied QDC scheme with distributed multiparticle entanglement.
Zhang {\it et al.} \cite{0010} have studied QDC scheme for
continuous variable. Hao {\it et al.} \cite{0011} and Fu {\it et
al.} \cite{0012} have proposed a controlled quantum dense coding
scheme by using a three-particle state and a four-particle state,
respectively. Liu {\it et al.} \cite{0013} have presented a QDC
protocol with a multi-level entangled state. Recently, Bru$\ss$
{\it et al.} \cite{0014} have presented a two-receiver quantum
dense coding protocol in a four-particle
Greenberger-Horne-Zeilinger (GHZ) state. However, in these
schemes, the quantum channels are symmetric, that is to say, the
dimension of the Hilbert space of the particle with sender is the
same as that of the particle with receiver.

On the other hand, Yan {\it et al.} \cite{0015} have given a QDC
scheme by using bipartite entangled state. Fan {\it et al.}
\cite{0016,0017} have given two QDC schemes by using the direct
product state. Fu {\it et al.} \cite{0018} have presented a QDC
scheme by using multipartite entangled state. In these schemes,
the quantum channels are non-symmetric, but the messages from the
sender can only be obtained by one receiver.

In this paper, we study a two-receiver quantum dense coding and an
$N$-receiver quantum dense coding in the case of non-symmetric
Hilbert spaces of the particles of the quantum channel.

Suppose Alice, Charlie, and Bob initially share a maximally
entangled state in $3\times3\times2\times2$-dimensional Hilbert
space in the following form:
\begin{equation}\label{1}
|\mu_{1}\rangle=\frac{1}{\sqrt{2}}\
(|0000\rangle+|1111\rangle)_{1234},
\end{equation}
where particles (1, 2) belong to Alice in $3^{2}$-dimensional
Hilbert space, particle 3 belongs to Charlie in 2-dimensional
Hilbert space, and particle 4 belongs to Bob in 2-dimensional
Hilbert space.

Firstly, Alice encodes one of her messages on qutrit 1 by
performing one of the six unitary transformations as follows:
\begin{equation}\label{2}
\begin{array}{ccc}
U_{00}=\left[%
\begin{array}{ccc}
  1\ &\ 0 &\ 0 \\
  0\ &\ 1 &\ 0 \\
  0\ &\ 0 &\ 1 \\
\end{array}%
\right],
U_{01}=\left[%
\begin{array}{ccc}
  1 & 0 & 0 \\
  0 & -1 & 0 \\
  0 & 0 & 1 \\
\end{array}%
\right],
U_{10}=\left[%
\begin{array}{ccc}
  0\ &\ 0 &\ 1 \\
  1\ &\ 0 &\ 0 \\
  0\ &\ 1 &\ 0 \\
\end{array}
\right],
\cr\\
U_{11}=\left[%
\begin{array}{ccc}
  0 & 0 & 1 \\
  1 & 0 & 0 \\
  0 & -1 & 0 \\
\end{array}%
\right],
U_{20}=\left[%
\begin{array}{ccc}
  0\ &\ 1 &\ 0 \\
  0\ &\ 0 &\ 1 \\
  1\ &\ 0 &\ 0 \\
\end{array}%
\right],
U_{21}=\left[%
\begin{array}{ccc}
  0 & -1 & 0 \\
  0 & 0 & 1 \\
  1 & 0 & 0 \\
\end{array}%
\right].
\end{array}
\end{equation}

Secondly, Alice encodes the rest of her messages on qutrit 2 by
performing one of the three unitary transformations $\{U_{00},
U_{10}, U_{20} \}$. Then, the collective unitary transformations
on qutrit 1 and qutrit 2 can be written as
\begin{equation}\label{3}
\begin{array}{ccc}
U^{+}_{1}=U_{00}\otimes U_{00},\ \ \ \ \ U^{-}_{1}=U_{01}\otimes
U_{00},\ \ \ \ \ U^{+}_{2}=U_{10}\otimes U_{00},
\cr\\
U^{-}_{2}=U_{11}\otimes U_{00},\ \ \ \ \ U^{+}_{3}=U_{20}\otimes
U_{00},\ \ \ \ \ U^{-}_{3}=U_{21}\otimes U_{00},
\cr\\
U^{+}_{4}=U_{00}\otimes U_{10},\ \ \ \ \ U^{-}_{4}=U_{01}\otimes
U_{10},\ \ \ \ \ U^{+}_{5}=U_{10}\otimes U_{10},
\cr\\
U^{-}_{5}=U_{11}\otimes U_{10},\ \ \ \ \ U^{+}_{6}=U_{20}\otimes
U_{10},\ \ \ \ \ U^{-}_{6}=U_{21}\otimes U_{10},
\cr\\
U^{+}_{7}=U_{00}\otimes U_{20},\ \ \ \ \ U^{-}_{7}=U_{01}\otimes
U_{20},\ \ \ \ \ U^{+}_{8}=U_{10}\otimes U_{20},
\cr\\
U^{-}_{8}=U_{11}\otimes U_{20},\ \ \ \ \ U^{+}_{9}=U_{20}\otimes
U_{20},\ \ \ \ \ U^{-}_{9}=U_{21}\otimes U_{20}.
\end{array}
\end{equation}
The above collective unitary transformations on qutrit 1 and
qutrit 2 will transform the state in Eq.~(\ref{1}) into the
corresponding state, respectively
\begin{equation}\label{4}
U^{\pm}_{1}|\mu_{1}\rangle=\frac{1}{\sqrt{2}}\
(|0000\rangle\pm|1111\rangle)_{1234}=|\mu^{\pm}_{1}\rangle,
\end{equation}
\begin{equation}\label{5}
U^{\pm}_{2}|\mu_{1}\rangle=\frac{1}{\sqrt{2}}\
(|1000\rangle\pm|2111\rangle)_{1234}=|\mu^{\pm}_{2}\rangle,
\end{equation}
\begin{equation}\label{6}
U^{\pm}_{3}|\mu_{1}\rangle=\frac{1}{\sqrt{2}}\
(|2000\rangle\pm|0111\rangle)_{1234}=|\mu^{\pm}_{3}\rangle,
\end{equation}
\begin{equation}\label{7}
U^{\pm}_{4}|\mu_{1}\rangle=\frac{1}{\sqrt{2}}\
(|0100\rangle\pm|1211\rangle)_{1234}=|\mu^{\pm}_{4}\rangle,
\end{equation}
\begin{equation}\label{8}
U^{\pm}_{5}|\mu_{1}\rangle=\frac{1}{\sqrt{2}}\
(|1100\rangle\pm|2211\rangle)_{1234}=|\mu^{\pm}_{5}\rangle,
\end{equation}
\begin{equation}\label{9}
U^{\pm}_{6}|\mu_{1}\rangle=\frac{1}{\sqrt{2}}\
(|2100\rangle\pm|0211\rangle)_{1234}=|\mu^{\pm}_{6}\rangle,
\end{equation}
\begin{equation}\label{10}
U^{\pm}_{7}|\mu_{1}\rangle=\frac{1}{\sqrt{2}}\
(|0200\rangle\pm|1011\rangle)_{1234}=|\mu^{\pm}_{7}\rangle,
\end{equation}
\begin{equation}\label{11}
U^{\pm}_{8}|\mu_{1}\rangle=\frac{1}{\sqrt{2}}\
(|1200\rangle\pm|2011\rangle)_{1234}=|\mu^{\pm}_{8}\rangle,
\end{equation}
\begin{equation}\label{12}
U^{\pm}_{9}|\mu_{1}\rangle=\frac{1}{\sqrt{2}}\
(|2200\rangle\pm|0011\rangle)_{1234}=|\mu^{\pm}_{9}\rangle.
\end{equation}
In the above equations, the notes $\pm$ correspond to the
superscripts for the collective unitary transformation and the
state composed of particles (1, 2, 3, 4). These 18 states above
are orthonormal:
$\langle\mu^{j}_{i}|\mu^{j'}_{i'}\rangle=\delta_{ii'}\delta_{jj'}$,
where $|\mu^{j}_{i}\rangle$ (or $|\mu^{j'}_{i'}\rangle$) is one of
the above 18 states in Eqs.~(\ref{4})$-$(\ref{12}); $i, i'$ $\in$
[0, 9]; $j, j'$ $\in$ $\{+, - \}$.

Thirdly, after performing one of these 18 collective unitary
transformations in Eq.~(\ref{3}), Alice sends her particle 1 to
Charlie and sends her particle 2 to Bob. Then, Charlie and Bob
share these 18 states in Eqs.~(\ref{4})$-$(\ref{12}) with equal
probabilities. In order to decode, Charlie makes a measurement
with the projectors
$P_{1}=|00\rangle\langle00|+|11\rangle\langle11|$,
$P_{2}=|10\rangle\langle10|+|21\rangle\langle21|$,
$P_{3}=|20\rangle\langle20|+|01\rangle\langle01|$, and
communicates the measurement result to Bob. If $P_{1}$ ($P_{2}$,
$P_{3}$) clicks, they know that the state they share is among the
three groups
$\{|\mu^{\pm}_{1}\rangle,|\mu^{\pm}_{4}\rangle,|\mu^{\pm}_{7}\rangle\}$
($\{|\mu^{\pm}_{2}\rangle,|\mu^{\pm}_{5}\rangle,|\mu^{\pm}_{8}\rangle\}$,
$\{|\mu^{\pm}_{3}\rangle,|\mu^{\pm}_{6}\rangle,|\mu^{\pm}_{9}\rangle\}$).
Now Bob performs a measurement with the same projectors $P_{1}$,
$P_{2}$, $P_{3}$, and communicates the measurement result to
Charlie. Depending on the outcomes of the projective measurements,
they know that the state they share is among which of the three
groups
$\{|\mu^{\pm}_{1}\rangle,|\mu^{\pm}_{4}\rangle,|\mu^{\pm}_{7}\rangle\}$
($\{|\mu^{\pm}_{2}\rangle,|\mu^{\pm}_{5}\rangle,|\mu^{\pm}_{8}\rangle\}$,
$\{|\mu^{\pm}_{3}\rangle,|\mu^{\pm}_{6}\rangle,|\mu^{\pm}_{9}\rangle\}$).

Note that none of the above projective measurements disturbs the
shared state. Lastly, depending on the outcomes of the previous
projective measurements, Charlie performs a measurement under the
basis $\{(|00\rangle\pm|11\rangle)_{13}\}$,
($\{(|10\rangle\pm|21\rangle)_{13}\}$,
$\{(|20\rangle\pm|01\rangle)_{13}\}$); Bob performs a measurement
under the basis $\{(|00\rangle\pm|11\rangle)_{24}\}$, or
$\{(|10\rangle\pm|21\rangle)_{24}\}$, or
$\{(|20\rangle\pm|01\rangle)_{24}\}$. Then, they communicate their
measurement results with each other, which will help them obtain
the messages from Alice. Because Charlie and bob communicate their
measurement results with each other, the amount of classical
information transmitted from Alice to Charlie and Bob is less than
$\log_{2}18$ bits. If the maximally entangled state in
Eq.~(\ref{1}) is shared between Alice and Charlie (Bob), where
particles (1, 2) belong to Alice in $3^{2}$-dimensional Hilbert
space, particles (3, 4) belong to Charlie (Bob) in
$2^{2}$-dimensional Hilbert space, the amount of classical
information transmitted from Alice to Charlie (Bob) is equal to
$\log_{2}18$ bits.

The above scheme can be directly generalized to the case that a
number of receivers are involved. The specific steps of the
process are depicted in FIG. 1. Let us follow the state through
this circuit. We assume that $(N+1)$-parties initially share a
maximally entangled state in $3^{N}\times2^{N}$-dimensional
Hilbert space in the following form:
\begin{equation}\label{13}
|\mu_{1}'\rangle=\frac{1}{\sqrt{2}}\
\left(\prod_{m=1}^{2N}|0\rangle_{m}+\prod_{m=1}^{2N}|1\rangle_{m}\right),
\end{equation}
where particles (1, 2, ..., $N$) belong to a sender in
$3^{N}$-dimensional Hilbert space, and each of the particles
$(N+1, N+2, ..., 2N)$ correspondingly belongs to each of the
receivers (1, 2, ..., $N$) in 2-dimensional Hilbert space. Now we
describe the scheme that a sender sends his messages to the
$N$-receivers by using the non-symmetric quantum channel in
Eq.~(\ref{13}). The scheme is composed of five steps.

(i) The sender encodes one of his messages on qutrit 1 by
performing one of the six unitary operations $\{U_{00}, U_{01},
U_{10}, U_{11}, U_{20}, U_{21}\}$.

\begin{figure}
\includegraphics{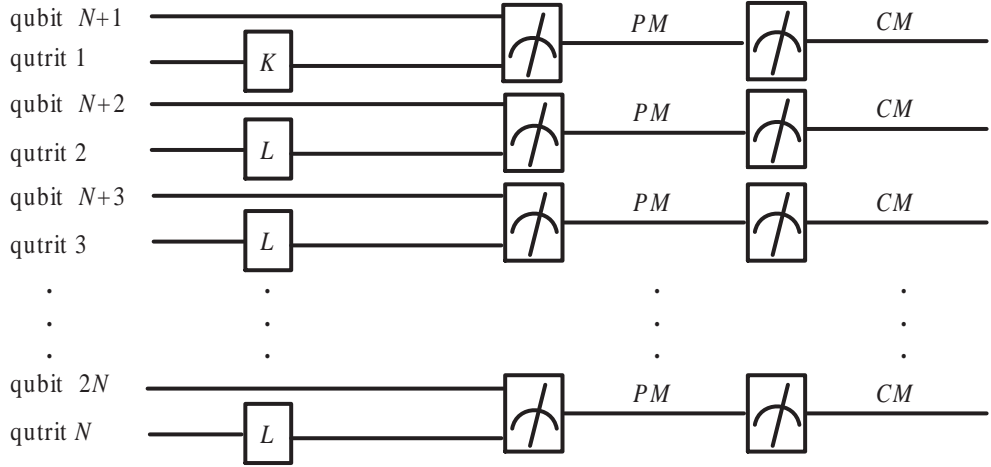}
\caption{Quantum circuit implementing $N$-receiver quantum dense
coding. $K$ $\in$ $\{U_{00}, U_{01}, U_{10}, U_{11},\\ U_{20},
U_{21}\}$, $L$ $\in$ $\{U_{00}, U_{10}, U_{20}\}$; $PM$ $\in$
$\{P_{1}, P_{2}, P_{3}\}$; $CM$ $\in$ $\{
(|00\rangle\pm|11\rangle), (|10\rangle\pm|21\rangle),
(|20\rangle\pm|01\rangle)\}$.}
\end{figure}

(ii) The sender encodes each of the rest of his messages on each
of the rest of his qutrits by performing one of the three unitary
operations $\{U_{00}, U_{10}, U_{20}\}$.

(iii) After performing his unitary operations, the sender sends
each of the particles (1, 2, ..., $N$) correspondingly to each of
the receivers (1, 2, ..., $N$).

(iv) In order to decode, the receiver 1, who receives the qutrit 1
from the sender, makes a measurement with the projectors
$P_{1}=|00\rangle\langle00|+|11\rangle\langle11|$,
$P_{2}=|10\rangle\langle10|+|21\rangle\langle21|$,
$P_{3}=|20\rangle\langle20|+|01\rangle\langle01|$; each of the
receivers (2, 3, ..., $N$) performs a similar measurement with the
same projectors $P_{1}$, $P_{2}$, $P_{3}$. Then, all the receivers
communicate their projective measurement results with one another.

(v) Finally, according to the outcomes of the previous projective
measurements, the receiver 1 performs a measurement under the
basis $\{ (|00\rangle\pm|11\rangle)_{1,1+ N}\}$,
($\{(|10\rangle\pm|21\rangle)_{1,1+N}\}$,
$\{(|20\rangle\pm|01\rangle)_{1,1+N}\}$); each of the receivers
(2, 3, ..., $N$) performs a similar measurement under the basis
$\{(|00\rangle\pm|11\rangle)_{ i,i+N}\}$, or
$\{(|10\rangle\pm|21\rangle)_{ i,i+N}\}$, or
$\{(|20\rangle\pm|01\rangle)_{ i,i+N}\}$, where $i$ $\in$ [2,
$N$]. Then, all the receivers communicate their measurement
results with one another, which will help them obtain the messages
from the sender.

In this way, the upper bound that the amount of classical
information transmitted from the sender to all the receivers by
using the non-symmetric quantum channel in Eq.~(\ref{13}) can be
expressed as
\begin{equation}\label{14}
C=\log_{2}(3^N \times2)= 1+N \log_{2}3.
\end{equation}

It must be stressed that our scheme is constructive. (i) We have
investigated a two-receiver quantum dense coding and generalized
it to an $N$-receiver quantum dense coding in the case of
non-symmetric Hilbert spaces of the particles of the quantum
channel. (ii) Compared with previous schemes, the sender can send
his messages to more receivers at the expense of some amount of
classical information in our scheme. (iii) Comparing the
two-receiver quantum dense coding with the one-receiver quantum
dense coding by the quantum channel in Eq.~(\ref{1}), the amount
of classical information transmitted in the one-receiver quantum
dense coding is the upper bound in the two-receiver quantum dense
coding. (iv) If and only if each of the receivers agrees to
collaborate, all the receivers can obtain the messages from the
sender. So each of the receivers may be as a controller during the
quantum dense coding. Obviously, our scheme can be applied to
quantum secret sharing and controlled quantum dense coding.

In conclusion, we have investigated a two-receiver quantum dense
coding and an $N$-receiver quantum dense coding in the case of
non-symmetric Hilbert spaces of the particles of the quantum
channel. The sender can send his messages to many receivers. The
scheme can be applied to quantum secret sharing and controlled
quantum dense coding.


\begin{thebibliography}{999}
\bibitem{0001}G. Ribordy, J. Brendel, J-D. Gautier, N. Gisin and H. Zbinden, Phys. Rev. A
{\bf 63}, 012309 (2001).
\bibitem{0002}J. G. Rarity, P. R. Tapster, P. M. Gorman and P. Knight, New J. Phys. {\bf
4}, 82 (2002).
\bibitem{0003}Y. Li, K. Zhang and K. Peng, Phys. Lett. A {\bf 324},
420 (2004).
\bibitem{0004}L. -Y. Hsu, Phys. Rev. A {\bf 68}, 022306 (2003).
\bibitem{0005}Z. J. Zhang, J. Yang, Z. X. Man and Y. Li, Eur. Phys. J. D
{\bf 33}, 133 (2005).
\bibitem{0006}T. Tyc, D. J. Rowe and B. C. Sanders, J. Phys. A {\bf 36}, 7625 (2003).
\bibitem{0007}C. H. Bennett and S. J. Wiesner, Phys. Rev.
Lett. {\bf 69}, 2881 (1992).
\bibitem{0008}H. J. Lee, D. Ahn and S. W. Hwang, Phys. Rev. A {\bf 66}, 024304
(2002).
\bibitem{0009}S. Bose, V. Vedral and P. L. Knight, Phys. Rev. A {\bf
57}, 822 (1998).
\bibitem{0010}J. Zhang and K. C. Peng, Phys. Rev. A {\bf 62},
064302 (2000).
\bibitem{0011}J. C. Hao, C. F. Li and G. C. Guo, Phys. Rev. A {\bf
63}, 054301 (2001).
\bibitem{0012}C. B. Fu, Y. Xia, B. X. Liu and S. Zhang, J. Korean Phys. Soc. {\bf
46}, 1080 (2005).
\bibitem{0013}X. S. Liu, G. L. Long, D. M. Tong and F. Li, Phys.
Rev. A {\bf 65}, 022304 (2002).
\bibitem{0014}D. Bru$\ss$, M. Lewenstein, A. Sen(De), U. Sen, G. M. D'Ariano and C. Macchiavello,
quant-ph/0507146.
\bibitem{0015}F. L. Yan and M. Y. Wang, Chin. Phys. Lett. {\bf 21}, 1195 (2004).
\bibitem{0016}Q. B. Fan, L. L. Sun, F. Y. Li, Z. Jin and S. Zhang, J. Korean Phys.
Soc. {\bf 21}, 769 (2005).
\bibitem{0017}Q. B. Fan and S. Zhang, Phys. Lett. A {\bf 348}, 160 (2006).
\bibitem{0018}C. B. Fu, Y. Xia and S. Zhang, Chin. Phys. (2006) in
press.
\end{thebibliography}
\end{document}